\documentclass[12pt,preprint]{aastex}

\begin{document}

\title{Flux-Limited Diffusion Approximation Models of Giant
Planet Formation by Disk Instability}

\author{Alan P.~Boss}
\affil{Department of Terrestrial Magnetism, Carnegie Institution of
Washington, 5241 Broad Branch Road, NW, Washington, DC 20015-1305}
\authoremail{boss@dtm.ciw.edu}

\begin{abstract}

 Both core accretion and disk instability appear to be required
as formation mechanisms in order to explain the entire range of 
giant planets found in extrasolar planetary systems. Disk instability
is based on the formation of clumps in a marginally-gravitationally 
unstable protoplanetary disk. These clumps can only be expected to contract 
and survive to become protoplanets if they are able to lose thermal energy 
through a combination of convection and radiative cooling. Here we 
present several new three dimensional, radiative hydrodynamics models
of self-gravitating protoplanetary disks, where radiative transfer is 
handled in the flux-limited diffusion approximation. We show that 
while the flux-limited models lead to higher midplane temperatures
than in a diffusion approximation model without 
the flux-limiter, the difference in temperatures does not appear
to be sufficiently high to have any significant effect on
the formation of self-gravitating clumps. Self-gravitating clumps
form rapidly in the models both with and without the flux-limiter.
These models suggest that the reason for the different outcomes of 
numerical models of disk instability by different groups cannot be 
attributed solely to the handling of radiative transfer, but rather 
appears to be caused by a range of numerical effects and assumptions.
Given the observational imperative to have disk instability form 
at least some extrasolar planets, these models imply that disk
instability remains as a viable giant planet formation mechanism.

\end{abstract}


\section{Introduction}

 Observations of protoplanetary disks around T Tauri stars traditionally 
imply disk masses in the range of 0.01 to 0.1 $M_\odot$ (Kitamura et al. 2002).
However, these disk masses may well be underestimated by as much as a factor
of 10 (Andrews \& Williams 2007). In addition, young stellar objects are 
likely to have had even higher disk masses at ages younger than typical 
T Tauri stars (a few Myr), as their protostellar disks transitioned 
into protoplanetary disks. Combined with the need for planet formation
theorists to prefer increasingly higher disk masses in order to
account for the timely formation of gas giants by core accretion (e.g.,
Inaba et al. 2003 suggest a 0.08 $M_\odot$ disk, while Alibert et al.
2005 considered disks as massive as 0.1 $M_\odot$), it is becoming
clear that at least some protoplanetary disks are likely to
have experienced a phase of gravitational instability, which might
have led to the rapid formation of gas giant planets by the disk
instability mechanism (e.g., Boss 1997; Mayer et al. 2002). The absence
of IR excesses in $\sim$ 65\% of the youngest stars observed by {\it Spitzer} 
suggests that the majority of protoplanetary disks dissipate on time scales
of $\sim$ 1 Myr or less (Cieza et al. 2007). While core accretion models can 
be constructed that permit giant planet formation times less than 1 Myr 
(Chambers 2006), other assumptions can require formation times of several 
Myr (e.g., Inaba et al. 2003; Alibert et al. 2005). 

 Considering that estimates of the frequency of gas giant planets 
around G dwarfs with orbits inside $\sim$ 20 AU range from $\sim$ 20\% 
to $\sim$ 40\%, there is a need for at least one robust formation 
mechanism for gas giant planets. It is important to note that {\it both}
core accretion and disk instability appear to be needed to explain 
the range of extrasolar planets detected to date. Core accretion would 
seem to be the preferred mechanism to form giant planets with very 
large inferred core masses. E.g., HD 149026b has been inferred to have a 
core mass of $\sim$ 70 $M_\oplus$ with a gaseous envelope of 
$\sim$ 40 $M_\oplus$ (Sato et al. 2005), though the formation of this 
planet is hard to explain even by core accretion (Ikoma et al. 2006).
Disk instability would seem to be the preferred mechanism for forming 
gas giants in very low metallicity systems (e.g., the M4 pulsar planet, 
where the metallicity [Fe/H] = -1.5 [Sigurdsson et al. 2003], and  
perhaps the giant planets orbiting HD 155358 and HD 47536, 
both of which have [Fe/H] = -0.68 [Cochran et al. 2007]). 
While disk instability is somewhat insensitive to metallicity
(Boss 2002), recent models have suggested that higher metallicity
could aid in the formation of giant planets by disk instability
(Mayer et al. 2007). Disk instability also appears to be needed to form 
gas giant planets around M dwarf stars (Boss 2006b), though the situation 
regarding formation by core accretion is unclear at present
(Laughlin et al. 2004; cf., Kornet et al. 2006).

 Based on all of these observations and detections, then, the 
main theoretical questions would seem to be whether there is 
a formation mechanism that can account for the {\it majority} 
of extrasolar planets, and if so, which mechanism it is. In
order to answer these questions, theorists have been busily examining
core accretion and disk instability in increasingly greater detail.
Core accretion has been subjected to considerably greater
scrutiny than disk instability, given that it has been the
generally accepted mechanism for giant planet formation for
almost three decades, dating back to Mizuno (1980), whereas
disk instability is only a decade-old as a serious alternative
to core accretion (Boss 1997), and is just now beginning to be
investigated sufficiently to discover its strengths and weaknesses.

 Theorists studying disk instability are divided into two distinct camps, 
those whose numerical models support the possibility of forming giant planets 
by this means (e.g., Boss 1997, 2000, 2001, 2002, 2004, 2005, 2006a,b, 2007; 
Mayer et al. 2002, 2004, 2007), and those whose 
numerical models (e.g., Pickett et al. 2000; Cai et al. 2006; 
Boley et al. 2006, 2007a,b) or analytical arguments (e.g., Rafikov 2007) 
do not. The reason for this basic difference in outcomes is presently unclear, 
and may be a combination of many effects (Nelson 2006; Boss 2007), such
as numerical spatial resolution, gravitational potential solver accuracy,
use of artificial viscosity, degree of stellar irradiation, detailed 
radiative transfer effects, and spurious numerical heating. 

 Recently attention has been focused on the role of radiative losses 
from the surface of the disk. A disk instability is likely to
be stifled if the optically-thick clumps that form are unable to 
lose at least some of the thermal energy produced by compressional
heating during contraction to protoplanetary densities. While vertical
convection appears to be able to cool the disk midplane (Boss 2004;
Boley et al. 2006; Mayer et al. 2007; Rafikov 2007), this thermal
energy must eventually be radiated away at the disk's surface.
Models employing the flux-limited diffusion approximation have
been presented by Boley et al. (2006, 2007b) and Mayer et al. (2007),
reaching opposite conclusions regarding the possibility of disk
instability forming protoplanets. The treatment of radiative 
boundary conditions for the disk differs for each group.
Boley et al. (2006) fit an atmosphere to the flux 
originating from the interior of the disk. Mayer et al. (2007) assume 
blackbody emission at the disk surface (for particles defined as being
on the surface), while the present models use an envelope bath
with a fixed temperature, typically 50 K. 

 With the exception of a single test model mentioned in passing by Boss (2001),
all of the author's disk instability models since Boss (2001) have employed
diffusion approximation radiative transfer {\it without} a flux-limiter,
for reasons of computational performance. We present here 
three new models that explore in some detail the effects of including
a flux-limiter in disk instability models, in the hopes of helping
to decide if this particular numerical choice is responsible for
the distinct disparity in outcomes of disk instability models.

\section{Numerical Methods}

 The calculations were performed with a code that solves the three 
dimensional equations of hydrodynamics and radiative transfer, as well
as the Poisson equation for the gravitational potential. This code 
has been used in all of the author's previous studies of disk instability, 
and is second-order-accurate in both space and time (Boss \& Myhill 1992). 

 The equations are solved on a spherical coordinate grid with $N_r = 101$,
$N_\theta = 23$ in $\pi/2 \ge \theta \ge 0$, and $N_\phi = 512$.
The radial grid is uniformly spaced with $\Delta r = 0.16$ AU between 4 
and 20 AU. The $\theta$ grid is compressed into the midplane to ensure 
adequate vertical resolution ($\Delta \theta = 0.3^o$ at the midplane).
The $\phi$ grid is uniformly spaced, and the central protostar is assumed
to move in such a way as to preserve the location of the center of 
mass of the entire system. The number of terms in the spherical harmonic 
expansion for the gravitational potential of the disk is $N_{Ylm} = 48$. 
The Jeans length criterion is monitored to ensure that numerical
artifacts do not form.
   
 The boundary conditions are chosen at both 4 and 20 AU to absorb radial
velocity perturbations. Mass and momentum that enters the innermost 
shell of cells at 4 AU are added to the central protostar and so removed 
from the hydrodynamical grid, whereas mass and momentum that reach the 
outermost shell of cells at 20 AU piles up at the boundary.

\section{Flux-Limited Diffusion Approximation}

 All of the author's disk instability models since Boss (2001) have
employed radiative transfer in the diffusion approximation, through
the solution of the equation determining the evolution of the 
specific internal energy $E$:

$${\partial (\rho E) \over \partial t} + \nabla \cdot (\rho E {\bf v}) =
- p \nabla \cdot {\bf v} + \nabla \cdot \bigl[ { 4 \over 3 \kappa \rho}
\nabla ( \sigma T^4 ) \bigr], $$

\noindent
where $\rho$ is the total gas and dust mass density, $t$ is time,
${\bf v}$ is the velocity of the gas and dust (considered to be a
single fluid), $p$ is the gas pressure, $\kappa$ is the Rosseland
mean opacity, $\sigma$ is the Stefan-Boltzmann constant, 
and $T$ is the gas and dust temperature. The energy 
equation is solved explicitly in conservation law form, as are the 
four other hydrodynamic equations.

 The final term in the energy equation represents the transfer of energy by
radiation in the diffusion approximation, which is valid in
optically thick regions of the disk. Given typical midplane
optical depths of $\sim 10^4$, the diffusion approximation should 
be valid at the disk midplane and throughout most of the disk, though
it will break down at the surface of the disk. In order to ensure that 
the diffusion approximation did not affect the solution in 
regions where it is not valid, Boss (2001) used a simple artifice
to control the flux in the low optical depth regions of the disk:
the divergence of the radiative flux term was set equal to zero
in regions wherever the optical depth $\tau$ dropped below a critical
value $\tau_{crit}$, where $\tau_{crit}$ was typically set equal to 10. 

 An alternative approach to treating the low optical depth regions
of disks in the diffusion approximation is to employ a flux-limiter
(e.g., Bodenheimer et al. 1990). The purpose of a flux-limiter is
to enforce the physical law that in low optical depth regions the ratio 
of the radiative flux ${\vec F}$ to the radiative energy density $e_r$ 
cannot exceed the speed of light $c$, i.e., $|{\vec F}| \le c e_r$.
Bodenheimer et al. (1990) adopted a prescription for enforcing
this constraint based on the flux-limiter proposed by Levermore \&
Pomraning (1981) for the situation where scattering of light is
negligible. The Levermore \& Pomraning (1981) flux-limiter is based on a 
heuristic argument leading to an approximation consisting of a rational 
function that uses a polynomial involving gradients of the radiation energy 
density. They then tested their formulation against an exact solution
for planar geometry, i.e., one-dimensional radiative transfer.
Their flux-limiter has been employed by Boley et al. (2006) and by
Mayer et al. (2007) in their disk instability calculations, with
differing results, as well as in the molecular cloud collapse models 
of Whitehouse \& Bate (2006). 

 The author's diffusion approximation code is derived from a
code that handles radiation transfer in the Eddington approximation
(Boss 1984; Boss \& Myhill 1992). In this code, the energy equation is
solved along with the mean intensity equation, given by

$$ {1 \over 3} {1 \over \kappa \rho} \nabla \cdot ( {1 \over \kappa \rho} 
\nabla J) - J = -B $$

\noindent
where $J$ is the mean intensity and $B$ is the Planck function
($B = \sigma T^4 / \pi$). The mean intensity $J$ is related to the 
radiative energy density $e_r$ by $J = c e_r / 4 \pi$, while the net flux
vector ${\vec H}$ is given by ${\vec H} = {\vec F} / 4 \pi$. Hence,
the statement of physical causality $|{\vec F}| \le c e_r$ is
equivalent to $|{\vec H}| \le J$. The Eddington approximation
version of the code does not calculate ${\vec H}$ directly,
but rather $\nabla \cdot {\vec H}$, as this quantity is used
in the code to calculate the time rate of change of energy per
unit volume due to radiative transfer, $L$, through

$$ L = - 4 \pi \nabla \cdot {\vec H} = 
   {4 \pi \over 3} \nabla \cdot ({1 \over \kappa \rho} \nabla J) $$

\noindent
in optically thick regions (Boss 1984). Hence, it is convenient
to apply the physical causality constraint $|{\vec H}| \le J$
in another form. Using the equation for $L$, one finds

$$ {\vec H} = - {1 \over 3 \kappa \rho} \nabla J. $$

\noindent
The constraint $|{\vec H}| \le J$ then becomes

$$ | {4 \pi \over 3 \kappa \rho} \nabla J | \le 4 \pi J. $$

\noindent
This constraint is then evaluated in a convenient but approximate 
manner by effectively taking the divergence of both sides of this
equation, resulting in a constraint on $L$ that

$$ |L| = |{4 \pi \over 3} \nabla \cdot ({1 \over \kappa \rho} \nabla J)| 
\le |4 \pi \nabla \cdot {\vec J}|,$$

\noindent
where ${\vec J}$ is a pseudovector with $J$ as components in all
three directions. In the diffusion approximation, $J = B$.
In practice, then, $L$ is calculated for each
numerical grid point, and if $|L|$ exceeds $|4 \pi \nabla \cdot {\vec J}|$,
$L$ is set equal to $|4 \pi \nabla \cdot {\vec J}|$ but with the
original sign of $L$ (i.e., preserving the sense of whether the
grid cell is gaining or losing energy through radiative transfer).

 Boss (2001) noted in passing that a model where this flux limiter
was employed did not result in any major changes in the progress
of the disk instability models under investigation, but provided
no details or justification for this statement. The main purpose 
of this paper is to return to this potentially key point, calculate
several new models with this version of flux-limited diffusion, and 
compare them to a disk instability model without a flux-limiter.

\section{Results}

 We now present the results of a set of three new models 
employing the flux-limiter defined in the previous section. 
The three models vary only in the value chosen for the critical
optical depth $\tau_{crit}$, below which the term calculating
the time rate of change of energy per unit volume due to radiative 
transfer, $L$ (effectively the divergence of the radiative flux), 
was set equal to zero. The three models employed $\tau_{crit} =$
0.1, 1.0, 10.0. In practice, all three of these models evolved
in very much the same manner, so figures will only be shown for the 
model with $\tau_{crit} = 1.0$, termed model FL1. The three models are
all continuations in time of model HR of Boss (2001), starting at 
a time of 322 yrs of evolution in model HR, and continuing for 
up to another 8 yrs of evolution ($\sim 1/2$ clump orbital period).
The new models are compared to model TE of Boss (2007), which 
used diffusion approximation radiative transfer, but without
the flux-limiter, and which also started from model HR of Boss (2001)
after 322 yrs of evolution.

 The results for models FL1 and TE at a time of 326 yrs of evolution
are shown in Figures 1 and 2. The midplane density contours for
the two models are very similar, especially so for the highest
density regions (cross-hatched). The densest spiral arms and
clumps that exist at this phase of the evolution are located
between 6 o'clock and 8 o'clock in Figures 1 and 2, with
maximum densities of $\sim 1.6 \times 10^{-9}$ g cm$^{-3}$ occurring
in the clumps at 6:30 o'clock. 

 Figures 3 and 4 depict the temperature and optical depth profiles as a 
function of vertical height above the midplane, along the $\theta$ 
coordinate direction, starting from the cells with the maximum 
densities in models FL1 and TE. Two different evaluations of the 
optical depth are plotted, namely the optical depth in
the radial direction (the value used in evaluating all
radiative transfer effects, including $L$ and $\tau_{crit}$, in
model TE and all previous models by the author, including the
flux-limiter test mentioned by Boss 2001), and the optical 
depth in the $\theta$ direction, which was used for evaluating 
$L$ and $\tau_{crit}$ in the three new models.
The decision of using an optical depth $\tau$
dependent only on the radial coordinate direction was originally made
in order to enforce consistency with spherically symmetric
calculations of protostellar cloud collapse, the problem that initially
motivated the development and testing of this radiative hydrodynamics
code (e.g., Boss \& Myhill 1992; Myhill \& Boss 1993). 
Figures 3 and 4 show that these two different evaluations
of $\tau$ do not differ greatly from each other, varying by no more
than a factor of 6 at the same vertical height. Given
the spatial resolution in the $\theta$ coordinate, the  
differences in the two evaluations of where $\tau = \tau_{crit}$ 
typically differ by less than one vertical cell. Improving this treatment of
$\tau$ in a three dimensional code may require the
use of an angle-dependent ray-tracing radiative transfer
routine, which would be prohibitively computationally expensive.
Alternatively, one could imagine using a weighted mean $\tau$ derived
from the values of $\tau$ in the three coordinate directions.
Nevertheless, it is apparent from Figures 3 and 4 that the
surface of the disk, defined as where $\tau \sim 2/3$, falls
at a vertical height of $\sim$ 1.6 to 1.7 AU in both models.

 Figures 5 and 6 are perhaps the most important for discerning
the effects of the flux-limiter, showing the vertical
temperature profiles over the maximum density clumps in
Figures 1 and 2 for models FL1 and TE. While the temperature
differences are hard to discern when plotted on a log scale
(Figure 3 and 4), on a linear scale it is clear that in
the flux-limiter model (FL1), there is a much steeper vertical
temperature gradient near the surface of the disk than in
the model without the flux-limiter (TE), as might be expected.
[Note that in both models, the temperature is assumed to fall
to 50 K in the disk's envelope (e.g., Chick \& Cassen 1997).]
In spite of this steeper rise below the disk's surface, however,
in both models FL1 and TE the profile flattens out near the
disk midplane and approaches essentially the same value of
$\approx 100$ K, with model FL1 being less than 2 K hotter than
model TE at the midplane. This similarity in thermal
behavior is consistent with the similarities in the density
evolution seen in Figure 1 and 2.

 The models assume that the disk is immersed in an envelope bath
at 50 K. The specific internal energy of the envelope gas in cells with 
densities less than $10^{-12}$ g cm$^{-3}$ is recalculated each time step
from the internal energy equation of state, using the assumed envelope 
temperature of 50 K and the envelope density at each grid cell. The
specific internal energy is thus forced to track the temperature
profiles displayed in Figures 5 and 6 and so to merge smoothly
with the assumed envelope thermal bath. This assumption can
lead to either the gain or the loss of internal energy, depending on
whether the envelope cell had a temperature lower than or greater
than 50 K before the envelope temperature constraint was applied. 
It is important to note that while the handling of the disk's
surface is directly linked to the ability of the disk to cool
itself by radiation into the infalling envelope, this surface treatment
has relatively little effect on the cooling of the midplane by
convective-like motions, as the driver for these motions is the
vertical temperature gradient near the disk's midplane, not the
disk's surface. Figure 5 in Boss (2004) shows that the
regions of convective instability according to the Schwarzschild
criterion are concentrated near the disk midplane, in spite of
the fact that the midplane is forced to be convectively stable
by the assumption of equatorial reflection symmetry. 

 Figure 7 and 8 display the results of both models after another 4 yrs 
of evolution, at $\sim$ 330 yrs, the maximum time to which model FL1 was 
evolved. It is evident again from these figures that the models continued
to evolve in a highly similar manner. In order to quantify this,
the dense clumps seen at 7:30 o'clock in Figures 7 and 8 were
evaluated in detail. For model FL1, this clump had a maximum
density of $1.2 \times 10^{-9}$ g cm$^{-3}$, and contained a
mass of 0.24 $M_{Jupiter}$ within regions with a density no less
than 1/30 of the maximum density. This mass exceeds the Jeans
mass of 0.23 $M_{Jupiter}$ for this clump, implying that it is
gravitationally bound. The ratio of thermal energy to gravitational
energy for the clump is 0.84. The equivalent spherical radius of the
clump was 0.38 AU, which is smaller than the critical tidal
radius of 0.49 AU, implying stability against tidal forces.
For comparison, the corresponding clump in model TE had a maximum
density of $1.5 \times 10^{-9}$ g cm$^{-3}$, containing a
mass of 0.30 $M_{Jupiter}$, compared to a Jeans mass of 0.24. 
The ratio of thermal energy to gravitational energy for this clump is 0.77. 
The equivalent spherical radius of this clump was 0.39 AU, 
also smaller than the critical tidal radius of 0.52 AU. While
model TE yielded a clump at this time that was 25\% more massive
than in model FL1, both models produced apparently 
self-gravitating clumps that could go on to form gas giant
protoplanets. The estimated orbital eccentricities and semimajor
axes are 0.033 and 11.3 AU for the clump in model FL1 and 0.004
and 11.3 for model TE clump at $\sim$ 330 yrs: both clumps are
on roughly circular orbits at this time.

 Evidently the clumps in both models are only marginally gravitationally
bound and marginally tidally stable, as shown by the fact that they 
tend to disappear within an orbital period or less. Calculations with
even higher spatial resolution have shown that the clumps become
better-defined as a result (Boss 2005), suggesting that in the
continuum limit, the clumps should survive to become protoplanets.
An adaptive mesh refinement code will be needed to properly investigate
the long-term survival of such clumps.  

 Figures 9 and 10 display the midplane temperature distributions
for models FL1 and TE at the same times as the density distributions
shown in Figures 7 and 8. The distributions are again highly
similar, at least in the outer disk and in the clump-forming
region. However, the region of the model FL1 disk inside about 6 AU
does appear to be considerably more nonaxisymmetric than
in the case of model TE, which is very nearly axisymmetric
inside 6 AU. Evidently use of the flux-limiter can lead to
significantly stronger nonaxisymmetric variations in the
temperature field. However, these temperature changes have
little effect on the clump-forming region of the disk, as the inner 
disk is the hottest region of the disk, with the midplane temperature
rising to over 630 K at the inner boundary at 4 AU, sufficiently
high to ensure gravitational stability ($Q >> 1$). Clumps
do not form in the inner region in these models because of the
high inner disk temperatures in the initial radial temperature
profile.

 In spite of the basic agreement after $\sim$ 8 yrs of evolution,
one must wonder what would happen on the much longer time
scales that must be considered in deciding whether these
clumps could survive to form gaseous protoplanets. In order
to address this question, Figures 11 and 12 show the time
evolutions of the volume-averaged midplane temperatures
and total midplane thermal energies for both models. The
intention is to discern if there are any trends evident
over 8 yrs of evolution that could be used to decide the
extent to which these two models might diverge if they
could be evolved arbitrarily farther in time. Figures 11 and 12
reveal no such evidence for divergence: both of these
quantities, when plotted for the entire midplane region
(Figure 11), or only for the region from 6.5 to 13 AU of
most interest for disk instability (Figure 12), show
that the two models evolve in very similar manners and
give no hint that their evolutions might turn out to
be significantly different if evolved even further in time.

 The models with the flux-limiter run considerably slower
than models without a flux-limiter, as in order to maintain a
stable solution of the energy equation with explicit time
differences, a smaller time step (often 1\% of the
Courant time step) had to be employed. This fact is evident
from Figures 11 and 12, which plot disk quantities every
10,000 time steps for models FL1 and TE: it is clear from the
density of plot symbols that model FL1 required many more time 
steps to evolve for the same period of time as model TE. The
flux-limiter models each required roughly one year of 
machine time on a dedicated Alpha workstation to run for only up to
8 yrs of model evolution time, i.e., the models were being
calculated only eight times faster than the disks were evolving in
model time, a situation similar to current weather prediction models.

 Rather than attempt to run these models significantly further in
time, then, one can address the question of the extent to
which the flux-limiter is having a long-term effect on the disk
by examining more closely the evolution of the innermost disk,
where the shorter orbital periods mean that the calculation 
has effectively been evolved for more dynamical times, i.e.,
for closer to a full orbital period. In order to be more
quantitative than is possible by presenting only density and
temperature contour plots, Figure 13 shows the amplitudes
of the $m = 1$ mode in the spherical harmonic representation
of the midplane density distribution, as a function of radial
distance, for models FL1 and TE. The time shown in Figure 13
was chosen in order to be as late as possible in the evolution
of model FL1, yet as close as possible in time to model TE (data 
files are only stored every 10,000 times steps, so the times 
available for cross-comparison are quite limited as a result.)
Figure 13 shows that the amount of nonaxisymmetry in the two 
models is nearly identical in the clump-forming region and beyond
(outer 2/3 in radius), but is still reasonably well-correlated
even in the innermost disk. At some radii, model FL1 has a higher
$m = 1$ amplitude than model TE, and the opposite is true at other 
inner disk radii. Figure 13 shows that the degree of nonaxisymmetry 
is well-correlated in both models, even in the innermost disk
where orbital periods are the shortest, suggesting that the
innermost disk shows little or no tendency for diverging
in behavior, at least over these time scales, as a result of the 
flux-limiter.

 Finally, Figures 14 and 15 show the effects of the flux-limiter
on the convective energy fluxes in models FL1 and TE at 
$\sim$ 330 yrs. The vertical convective energy flux is calculated
as in Boss (2004) as the product of the local vertical
velocity, cell area, specific internal energy, and cell density.
Figures 14 and 15 plot this flux for a conical surface at a 
fixed angle of 0.3 degrees above the disk's midplane (i.e.,
the $J = 2$ cells in the $\theta$ coordinate).
The convective flux must vanish at the midplane as a consequence 
of the assumed equatorial symmetry of the models; if this
constraint were to be lifted, more vigorous convective fluxes are
to be expected (e.g., Ruden et al. 1988). These two figures
show that application of the flux-limiter has no obvious
systematic effects on the vertical convective fluxes near
the midplane, where the need for convective cooling is
most severe; the overall patterns of upwelling and downwelling
regions are quite similar in both models.

\section{Discussion}

 These models have shown that the flux-limiter has relatively
little effect on the evolution, at least during a phase when the
disk has already begun forming strong spiral arms and clumps. 
The question arises as to what would happen if the flux-limiter
was applied earlier in the evolution of the disk, prior to
the formation of highly nonaxisymmetric structures. The model
noted in passing by Boss (2001) began at an earlier time than
in the present models, after 141 yrs of evolution instead of
after 322 yrs, and so tested the effects of the flux-limiter
at such an earlier phase. Unfortunately, the data files from 
the Boss (2001) flux-limiter model no longer exist, as
the models were run in 2000 and stored on a hard disk that
has since failed. Hence it is not possible to present those
results in the detail presented here, a fact that motivated
calculation of the models in this paper. The flux-limiter
model from Boss (2001) was compared to a non-flux-limiter
model by visual inspection of density contour plots, as
in Figures 1 and 2 and in Figures 7 and 8 in the present paper,
with the conclusion being that there were no significant
differences apparent in the degree of clumpiness in the two
models. While purely a qualitative judgement, these results
suggest that the role of the flux-limiter is similarly
limited both early and late in the development of a phase
of disk instability.

 Boley et al. (2006, 2007b) have presented the results of a
series of tests of their radiative hydrodynamics code on a 
``toy problem'' (the plane-parallel grey atmosphere)
with sufficient assumptions to permit an 
analytical solution for the temperature distribution. Their
toy problem assumes an infinite slab, making the problem the
same as a one-dimensional Cartesian atmosphere. This toy problem 
is well-suited to their cylindrical coordinate code, as their vertical 
($z$) cylindrical coordinate is effectively a one-dimensional Cartesian 
coordinate, and by freezing motion in the radial direction
(Boley et al. 2006) and applying suitable boundary conditions
at the disk edges the Boley et al. code can be used to simulate
a plane-parallel atmosphere. 

 Boley et al. (2006) ``... challenge
all researchers who publish radiative hydrodynamics simulations
to perform similar tests or to develop tests of their own and
publish the results.'' While it would be ideal to be able to undertake 
the same tests as those examined by Boley et al. (2006, 2007b), the fact
that their tests assume a plane-parallel atmosphere makes them
unsuitable for a spherical coordinate code, which has no Cartesian
coordinate. The closest analogue coordinate for the present code would be the 
$\theta$ coordinate, but trying to reproduce a plane-parallel atmosphere 
solution with spherical coordinates places the spherical coordinate code
at a distinct disadvantage from the beginning, as any attempt
to study a plane-parallel atmosphere with such a code will
immediately introduce corrugations in all variables in each 
azimuthal ($r$, $\theta$) plane. One could perhaps average over
the entire disk to try to remove these corrugations, but the non-uniform
$\theta$ grid spacing, designed to represent realistic protoplanetary
disks, not plane-parallel slabs, would result in highly variable
effective spacings in the vertical direction, which would further
complicate the analysis. Studying the performance of the current 
radiative hydrodynamics code on the Boley et al. (2006, 2007b)
tests in an unbiased manner requires writing a new one dimensional
radiative hydrodynamics code based on the same numerical assumptions
as the present three dimensional spherical coordinate code.
Writing and testing such a code, even before trying the Boley et 
al. (2006, 2007b) tests, is a non-trivial task, as no such code
exists. Writing such a code to perform the Boley et al. (2006, 2007b)
tests would be a worthy goal for future work.

 Alternatively, it is possible that a spherical coordinate
version of the Boley et al. (2006, 2007b) tests could be posed
and examined with the one dimensional spherical coordinate version
of the present code. This would also meet the request by Boley
et al. (2006) ``... to develop tests of their own and publish 
the results.'' This task remains for future investigation.

 Finally, it should be noted that the motivation of this 
paper is the same as that expressed in the Boley et al. (2006) 
request ``... to develop tests of their own and
publish the results.'' Many other numerical tests of the present
code have been presented as follows: spatial resolution
(Boss 2000, 2005); gravitational potential solver (Boss 2000,
2001, 2005), artificial viscosity (Boss 2006a); and 
radiative transfer (Boss 2001, 2007). It would be valuable
for other reseachers to consider their own tests of all
of these key numerical aspects.

\section{Conclusions}

 The results presented here confirm the statement made by Boss (2001)
that the inclusion of a flux-limiter in these calculations does
not lead to significantly different outcomes for the progress
of a disk instability calculation. Even with the
steeper vertical temperature gradient near the disk surface when
a flux-limiter was employed (Figures 7 and 8), the corresponding
midplane temperature increased by no more than 2\%. Similarly,
Boss (2007) investigated the effects of several other changes 
in the treatment of radiative transfer in these models,
finding that the numerical assumption that had the largest
effect was the relaxation of the monotonically declining vertical
temperature profile, which resulted in clumps that were no more
than a factor of 2 times less dense than when monotonicity was enforced.
For comparison, for models FL1 and TE in Figure 7 and 8, the
maximum clump densities differed by only 25\%, implying even less
of a difference between models FL1 and TE and the two models (H and TZ)
from Boss (2007). 

 Evidently disk instability is tolerant of a range of treatments
of the radiative transfer, at least up to a point. If there
is a means for a clump to cool enough to contract, the clump
will find this means to allow its survival. In this context it
is of interest to note that analytical evaluations of disk instability 
(e.g., Rafikov 2007) have been restricted to considering plane
parallel (one dimensional) disk models, where the entire
disk midplane must be cooled, in order to cool the disk
midplane anywhere at all. In a more realistic three dimensional
disk model, of the sort depicted in the present numerical models,
only the limited midplane region inside the dense clump needs to lose
thermal energy, in any direction, in order for the clump
to continue to contract and possibly survive to become
a gas giant planet. This is a considerably relaxed criterion
for cooling and ultimate clump survival compared to the cooling of an 
entire slab of midplane gas and dust. Similarly, a hot spot
on the disk surface above a contracting clump will find it
easier to radiate away its thermal energy than if the entire
disk surface has the same vertical thermal profile as that under
the hot spot.

 Given the apparent observational need for disk instability to be
able to form gas giant planets in some protostellar environments,
if flux-limiters and other radiative transfer effects are not
the main reason for the discrepant outcomes in models of disk
instability, then there must be other reasons, or combinations
of reasons, for these differences, as examined and discussed
in some detail by Nelson (2006) and Boss (2007). Spurious
heating of the inner disk associated with numerical oscillations 
is one possible source of these discrepancies that deserves 
further scrutiny (Boley et al. 2006, 2007b), as this leads to gravitational 
stability in the same region of the disk where clumps form in other
disk instability models (Boss 2007).

 Because of the unsatisfactory nature of the theoretical understanding 
of disk instabilities at present, it is important to continue
to undertake code tests. The present models have shown that
the use of a flux-limiter has relatively little effect on the
evolution of an instability during the phase when the disk is
already dynamically unstable. However, it is also important
to investigate the role of a flux-limiter during earlier phases
of evolution, before the disk becomes unstable, in order to learn
if the flux-limiter can affect clump formation if applied from
the very beginning of the evolution. A new model is underway that
investigates this possibility, and the results will be presented
in a future paper. Other code tests should also be sought, similar
to the radiative transfer tests advanced by Boley et al. (2006,
2007b), except for spherical geometry instead of slab geometry,
so that the present code can be tested in a similar manner. 

\acknowledgements

 I thank the referee for a number of good ideas for improving the manuscript,
and Sandy Keiser for computer systems support. This research was 
supported in part by NASA Planetary Geology and Geophysics grants NNG05GH30G 
and NNX07AP46G, and is contributed in part to NASA Astrobiology Institute
grant NCC2-1056. The calculations were performed on the Carnegie Alpha
Cluster, the purchase of which was partially supported by NSF Major Research
Instrumentation grant MRI-9976645.

\clearpage

\begin{figure}
\vspace{-2.0in}
\plotone{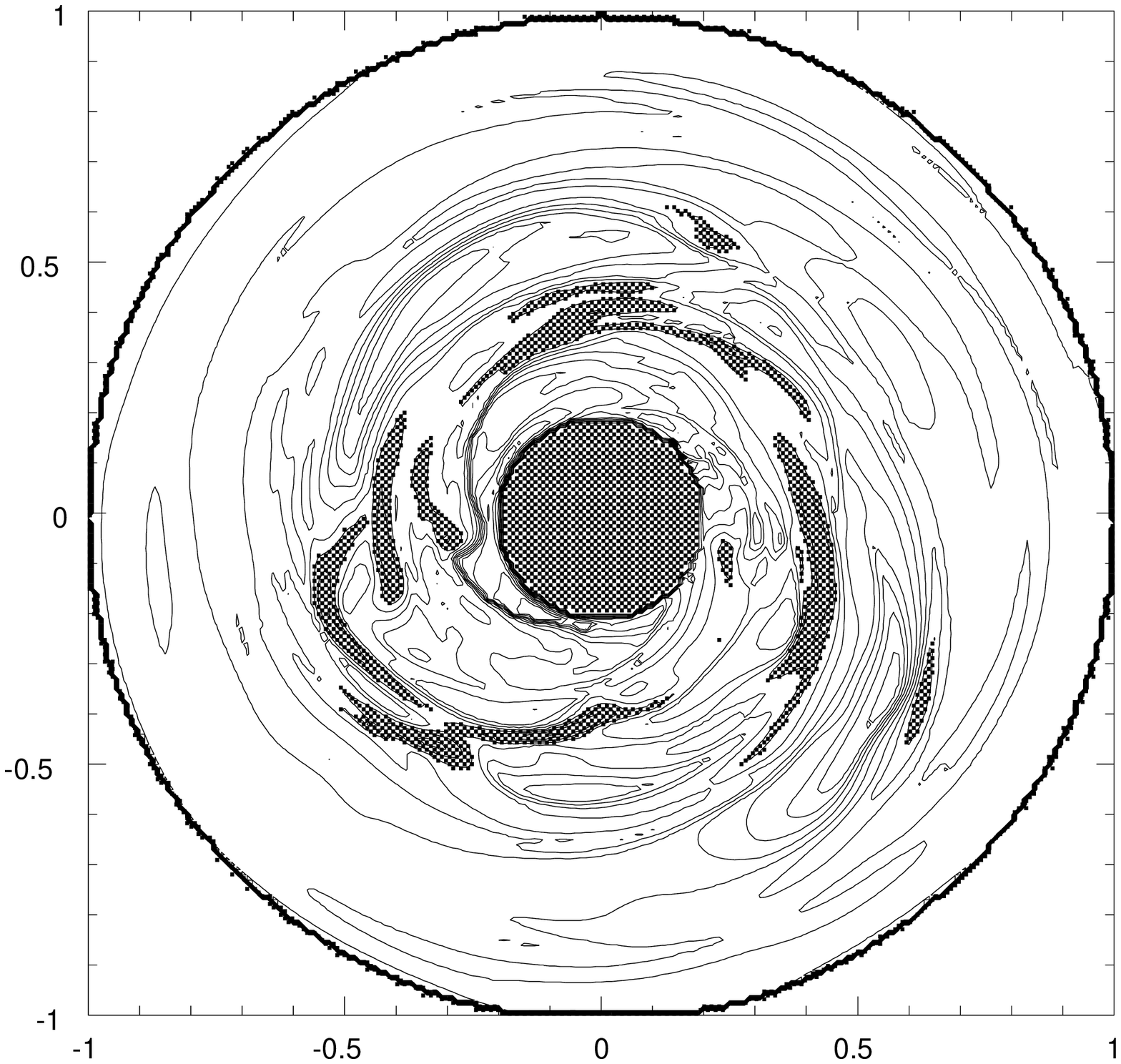}
\caption{Equatorial density contours for model F1 after 325.8 yrs of evolution.
The disk has an outer radius of 20 AU and an inner radius of 4 AU.
Hashed regions denote clumps and spiral arms with densities higher than
$10^{-10}$ g cm$^{-3}$. Density contours represent factors of two
change in density.}
\end{figure}

\clearpage

\begin{figure}
\vspace{-2.0in}
\plotone{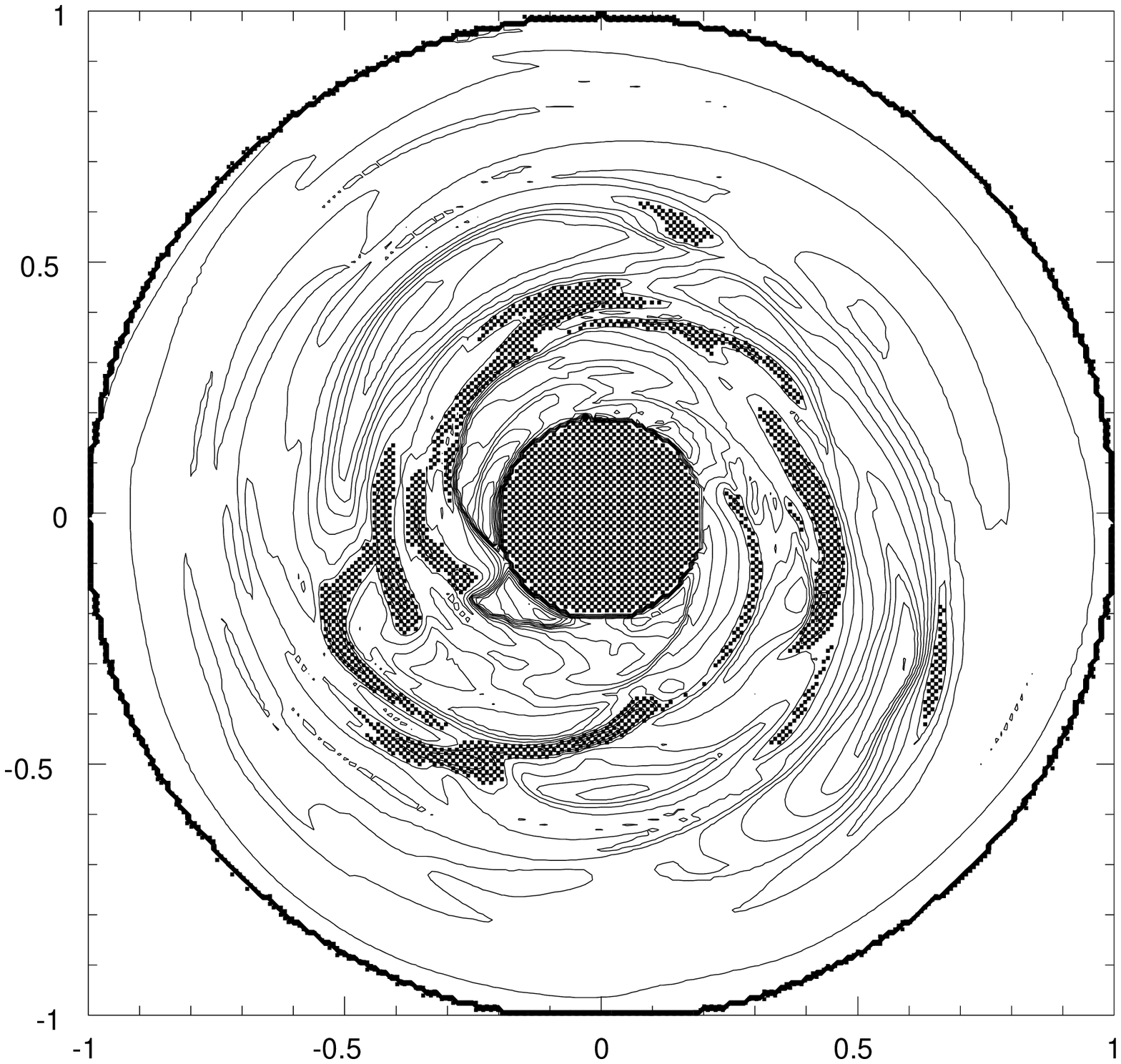}
\caption{Same as Figure 1, but for model TE after 326.5 yrs.}
\end{figure}

\clearpage

\begin{figure}
\vspace{-2.0in}
\plotone{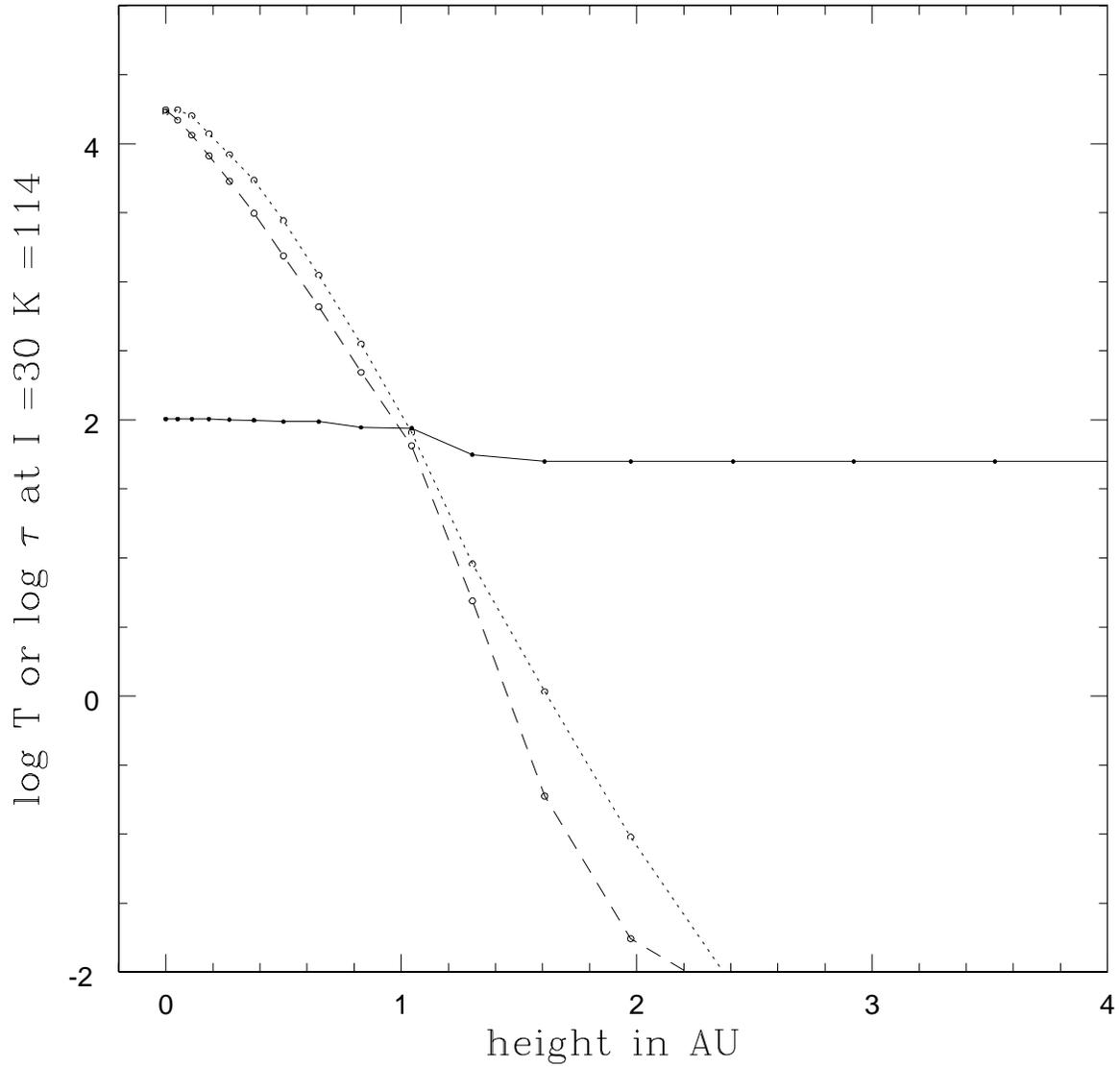}
\caption{Log of the optical depth (dashed lines) and 
temperature (solid line) as a function of distance above
the midplane for model FL1 after 325.8 yrs of evolution. The profiles
are along the $\theta$ coordinate, starting at the location
of the maximum density in the midplane. The long-dashed line gives
the optical depth in the $\theta$ direction, starting with zero
at the rotational (symmetry) axis, while the short-dashed line
gives the optical depth in the radial direction, starting with
zero at the outer edge of the spherical computational volume.}
\end{figure}

\clearpage

\begin{figure}
\vspace{-2.0in}
\plotone{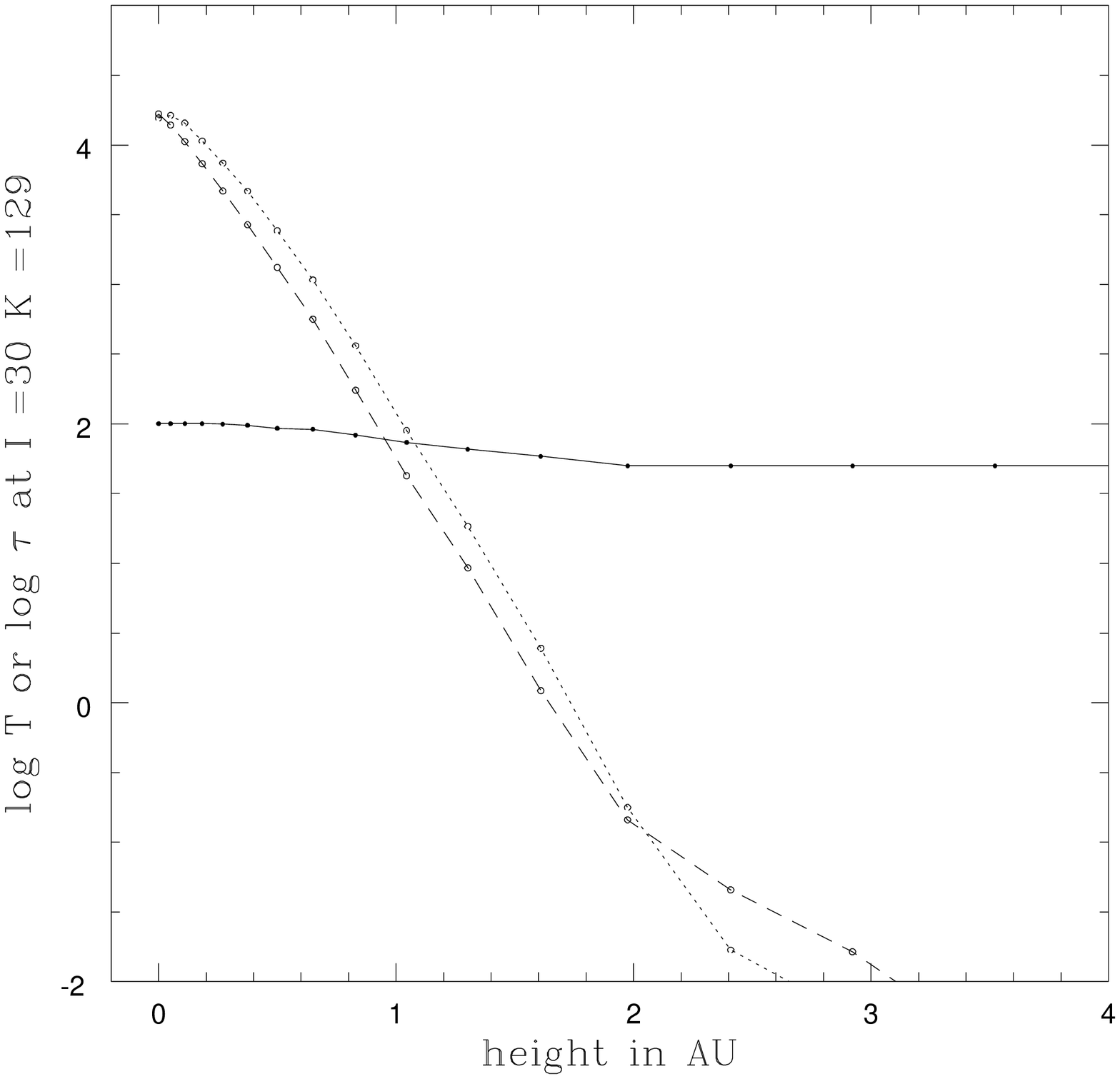}
\caption{Same as Figure 3, but for model TE after 326.5 yrs.}
\end{figure}

\clearpage

\begin{figure}
\vspace{-2.0in}
\plotone{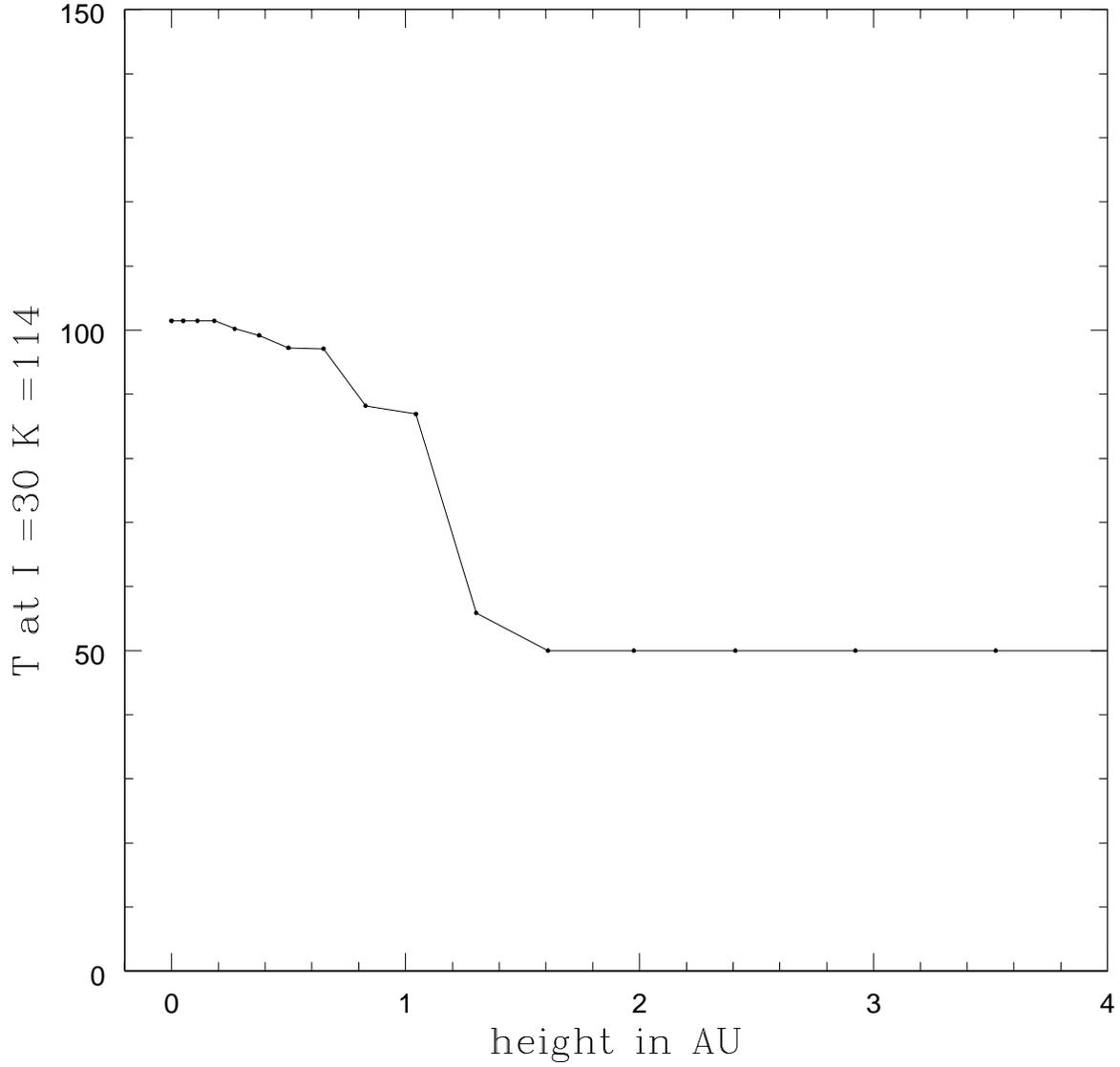}
\caption{Temperature (linear scale) as a function of distance above the 
midplane for model FL1 after 325.8 yrs of evolution, plotted as in Figure 3.}
\end{figure}

\clearpage

\begin{figure}
\vspace{-2.0in}
\plotone{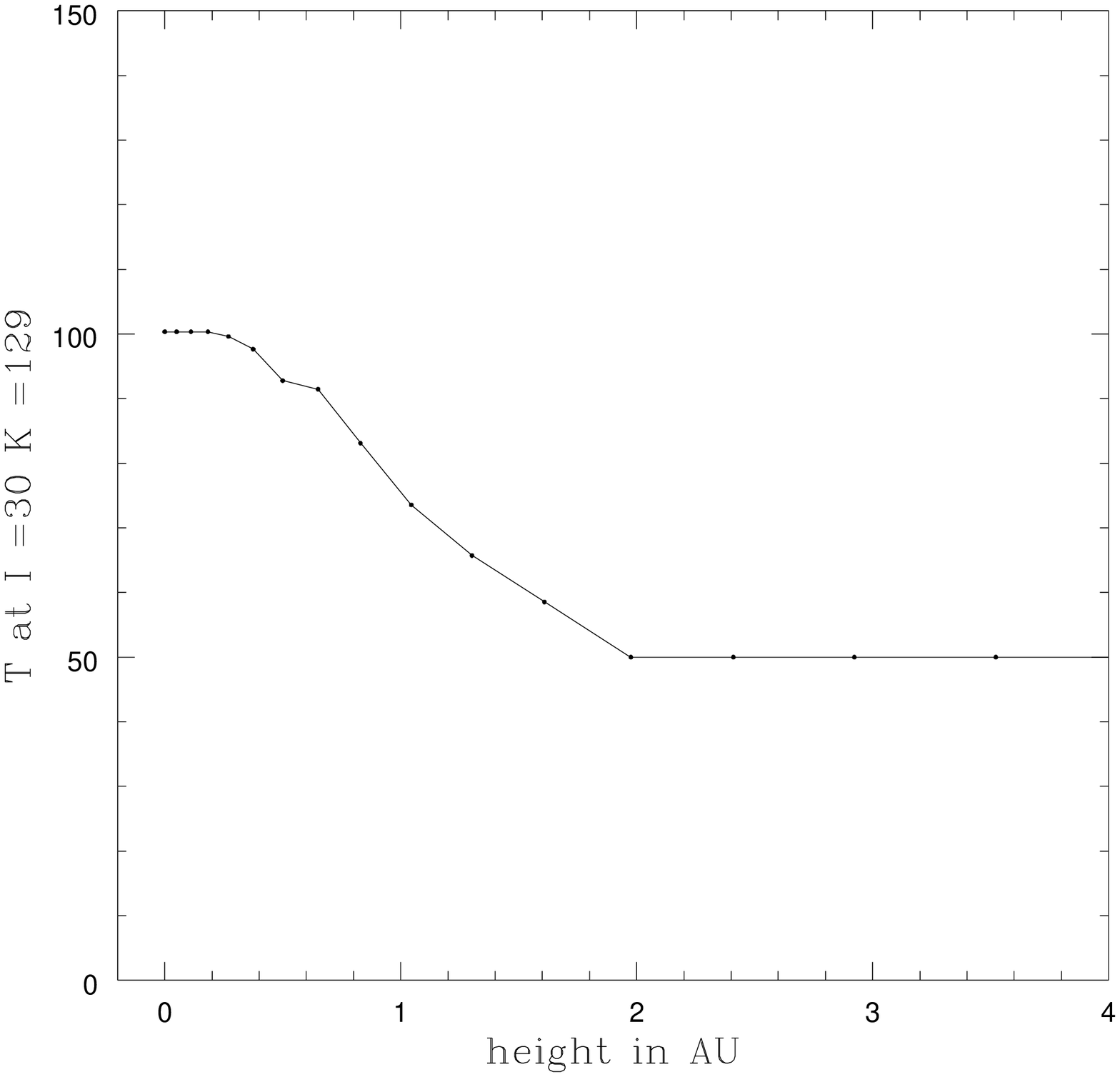}
\caption{Same as Figure 5, but for model TE after 326.5 yrs.}
\end{figure}

\clearpage

\begin{figure}
\vspace{-2.0in}
\plotone{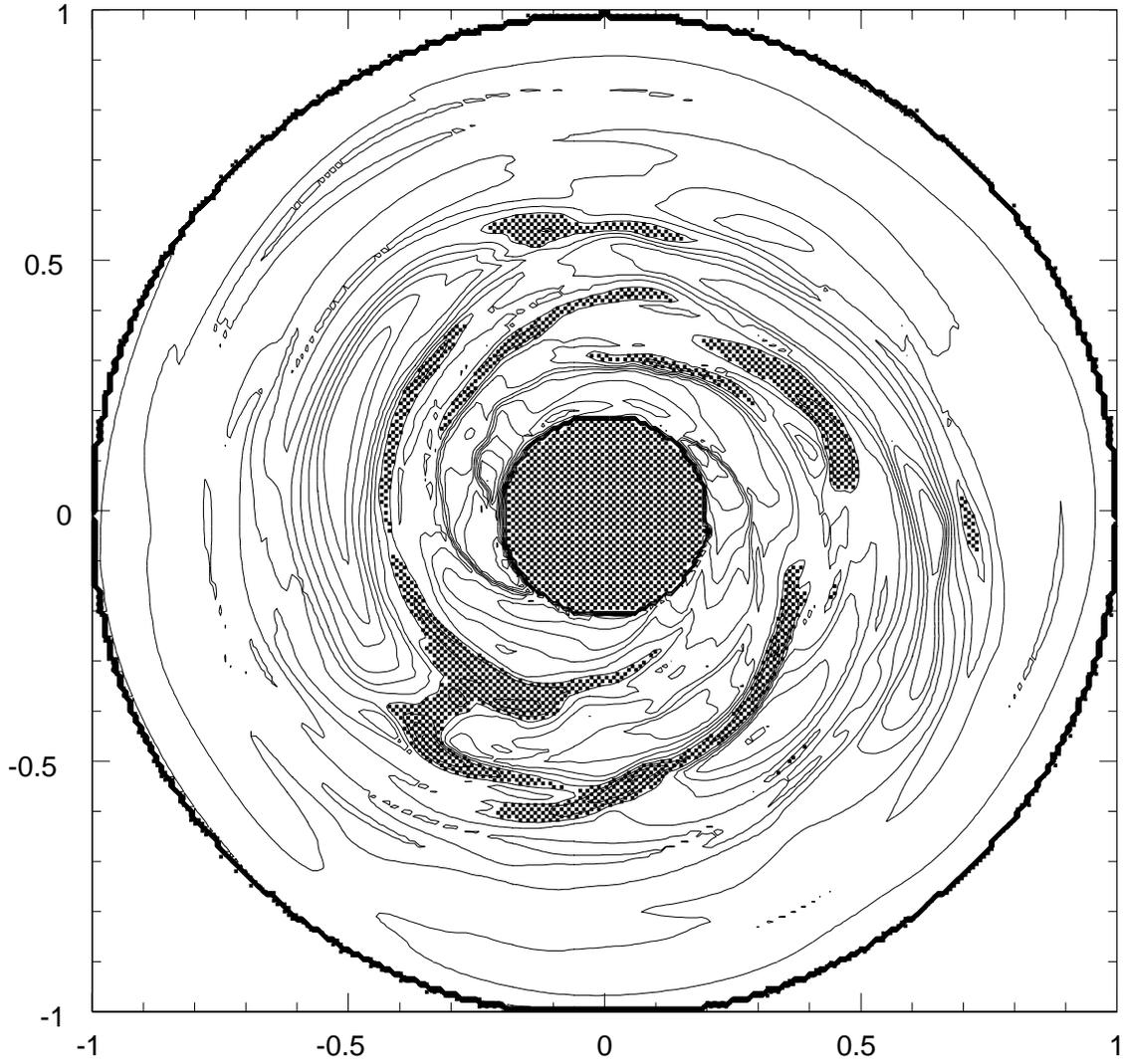}
\caption{Equatorial density contours for model FL1 after 329.6 yrs of 
evolution, plotted as in Figure 1.}
\end{figure}

\clearpage

\begin{figure}
\vspace{-2.0in}
\plotone{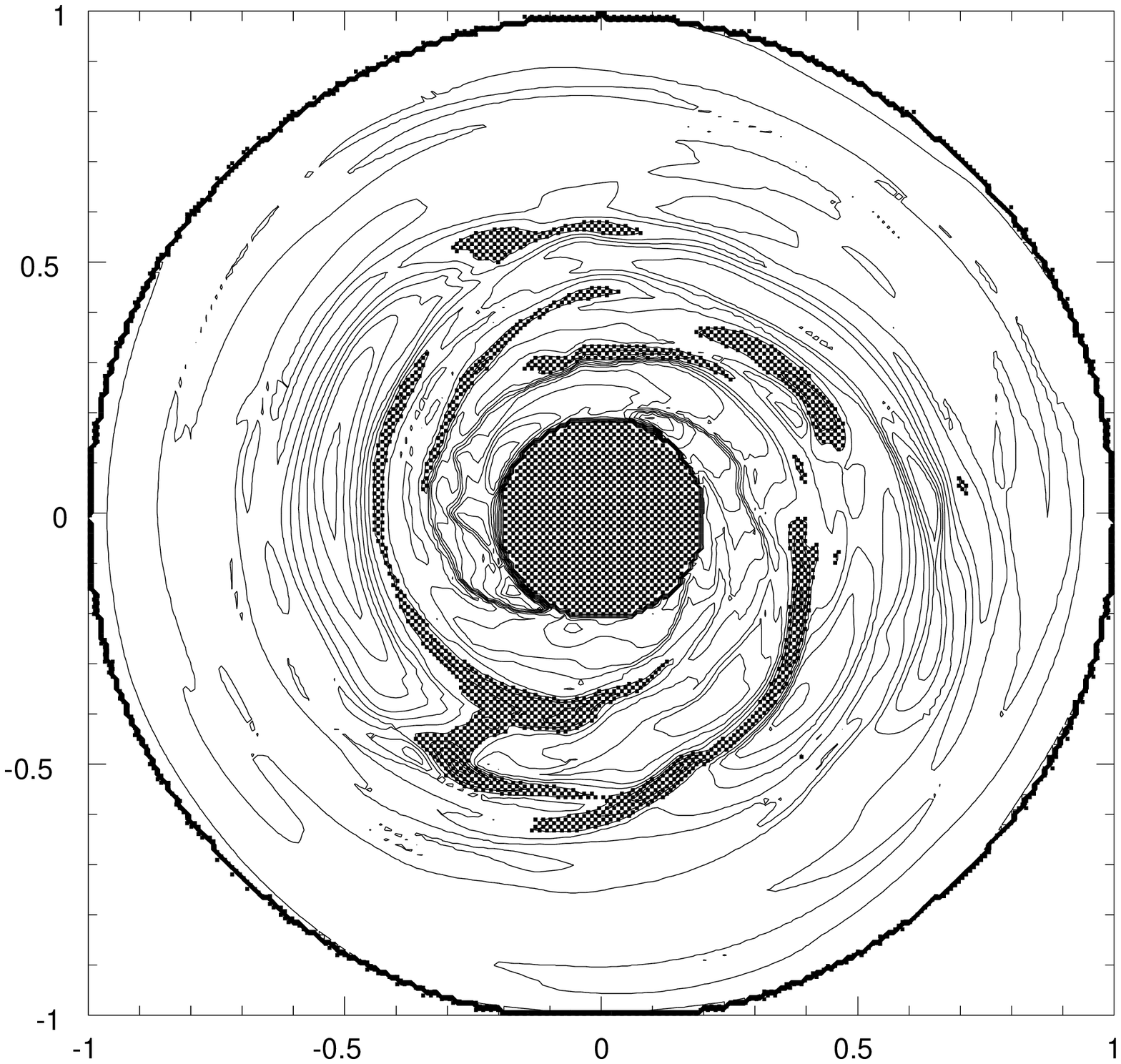}
\caption{Same as Figure 7, but for model TE after 330.3 yrs.}
\end{figure}

\clearpage

\begin{figure}
\vspace{-2.0in}
\plotone{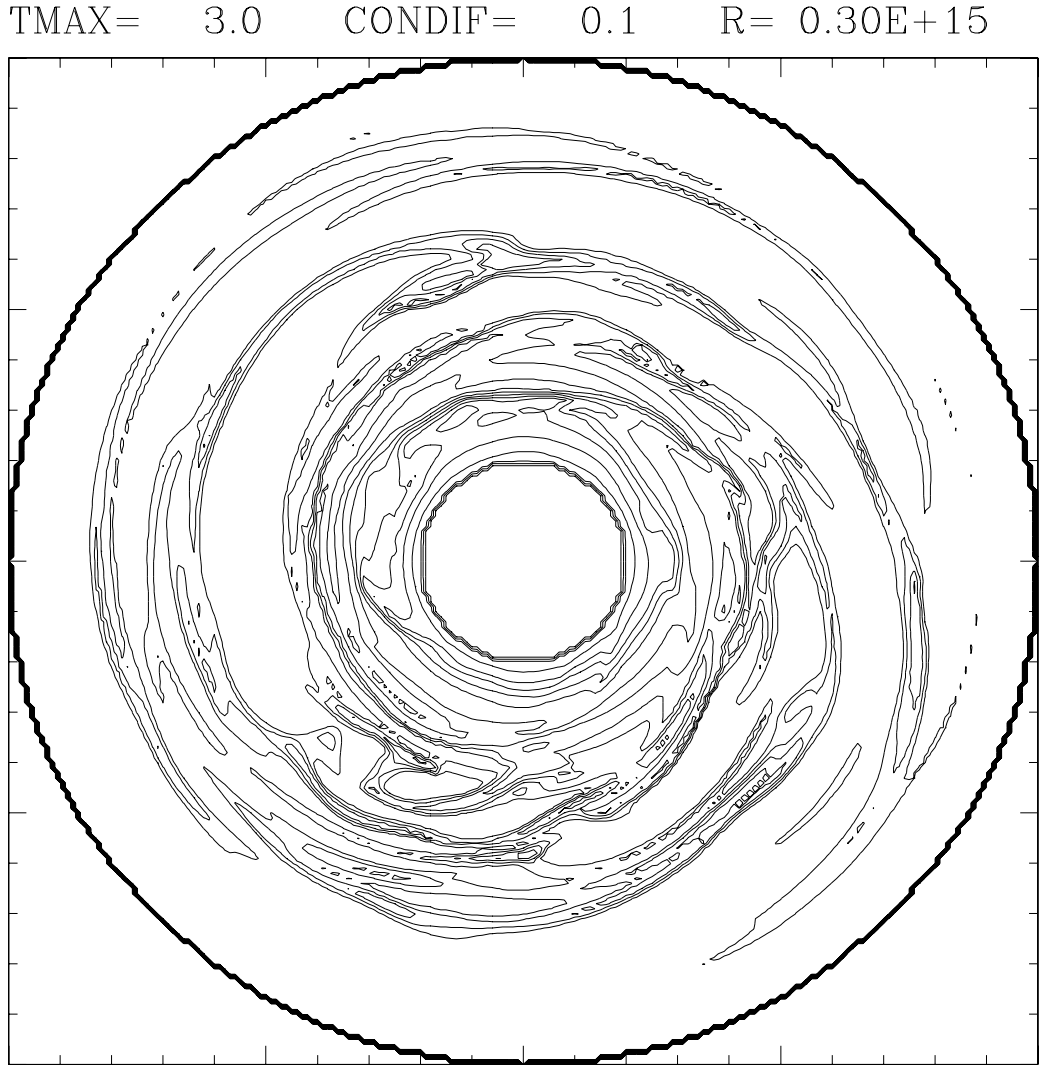}
\caption{Equatorial temperature contours for model FL1 after 329.6 yrs of 
evolution, plotted as in Figure 1, with temperature contours representing 
factors of 1.26 changes in temperature.}
\end{figure}

\clearpage

\begin{figure}
\vspace{-2.0in}
\plotone{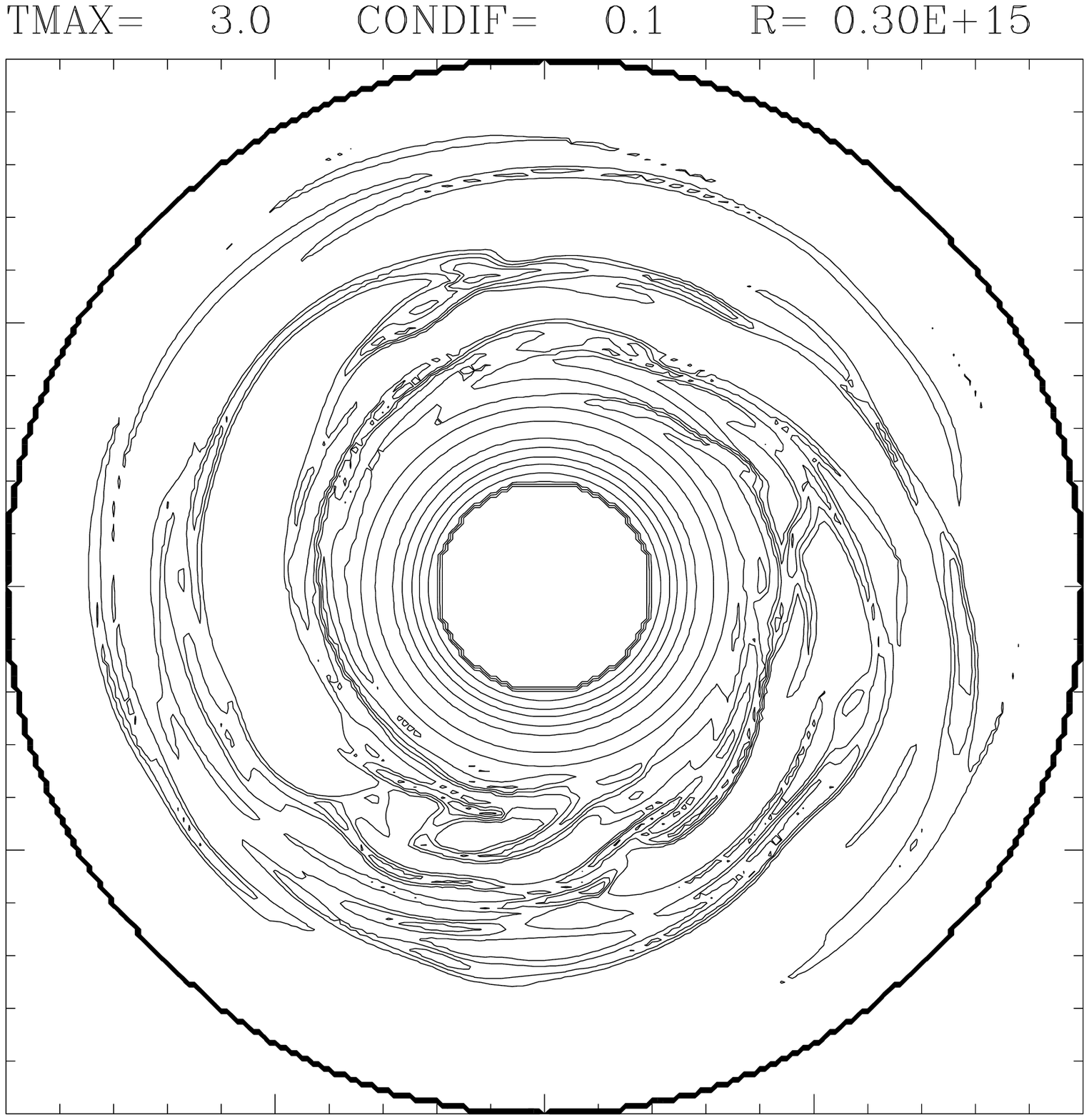}
\caption{Same as Figure 9, but for model TE after 330.3 yrs.}
\end{figure}

\begin{figure}
\vspace{-2.0in}
\plotone{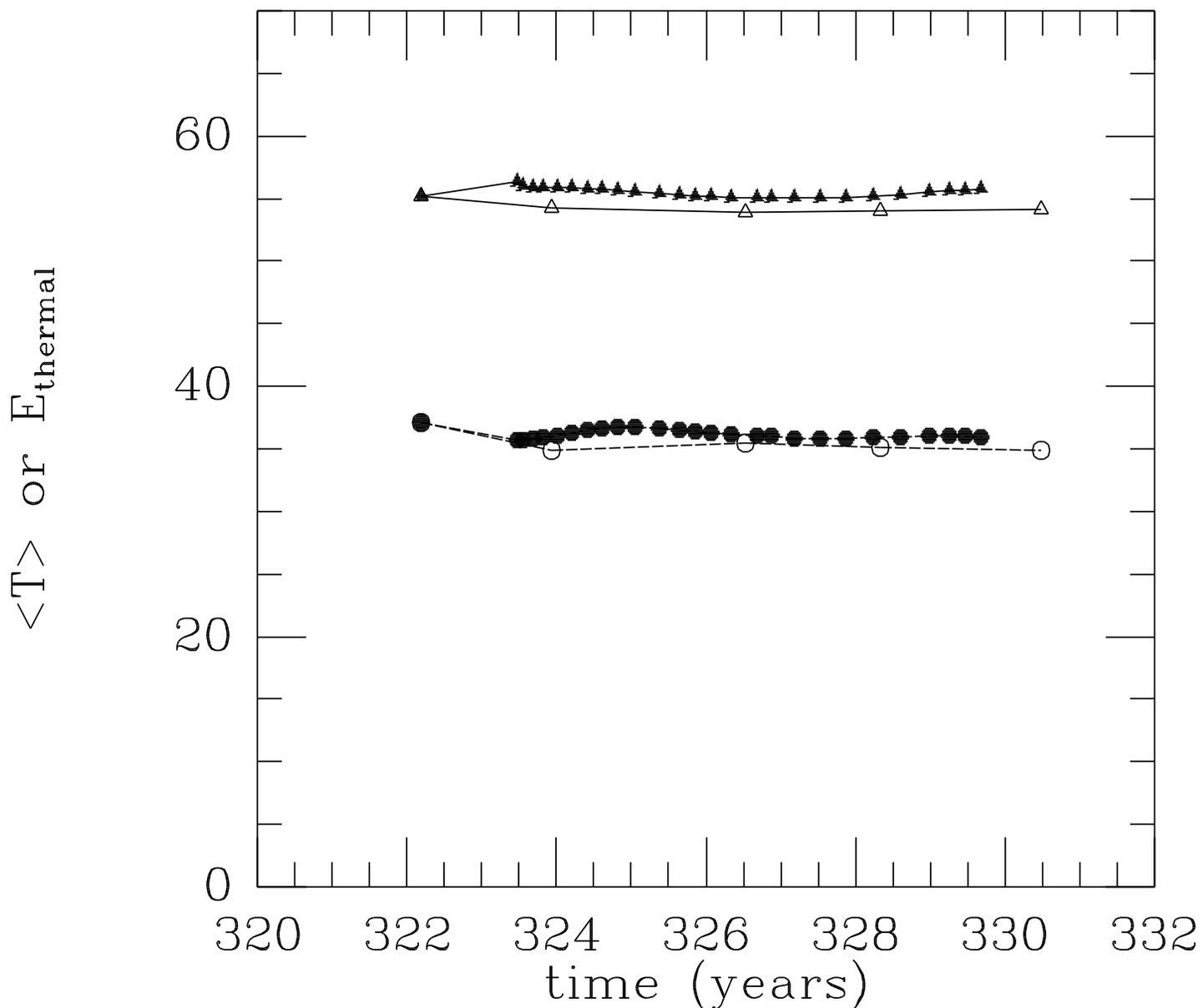}
\caption{Volume-averaged midplane temperatures (triangles and solid
lines) and total midplane $E_{thermal}$ (circles and dashed lines)
for model FL1 (filled symbols) and model TE (open symbols) as a
function of time in years. Temperatures are given in K and the total thermal
energy in units of $10^{39}$ ergs.}
\end{figure}

\clearpage

\begin{figure}
\vspace{-2.0in}
\plotone{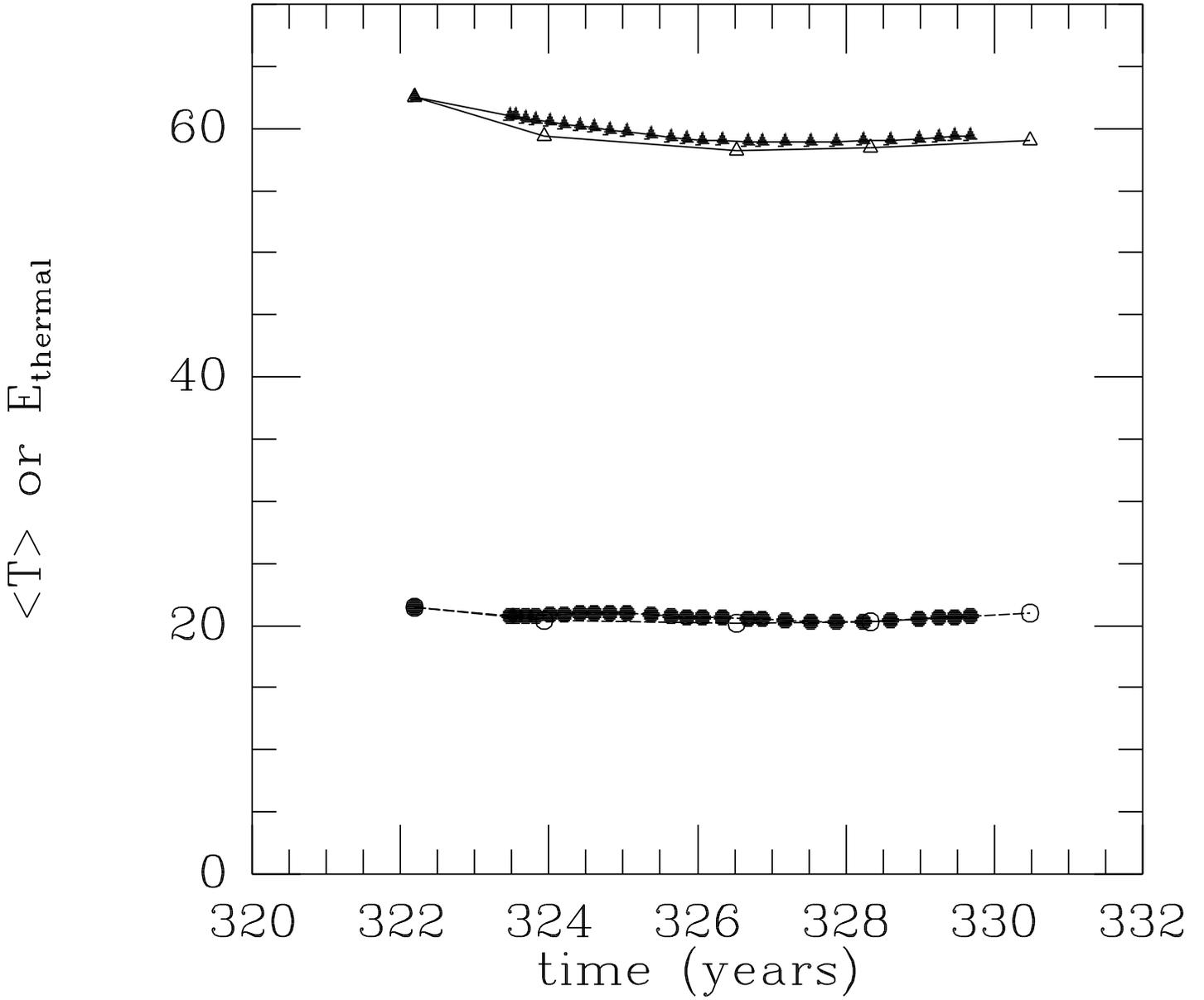}
\caption{Same as Figure 11, but only for radial distances of
6.5 to 13 AU in the midplane.}
\end{figure}

\clearpage

\begin{figure}
\vspace{-2.0in}
\plotone{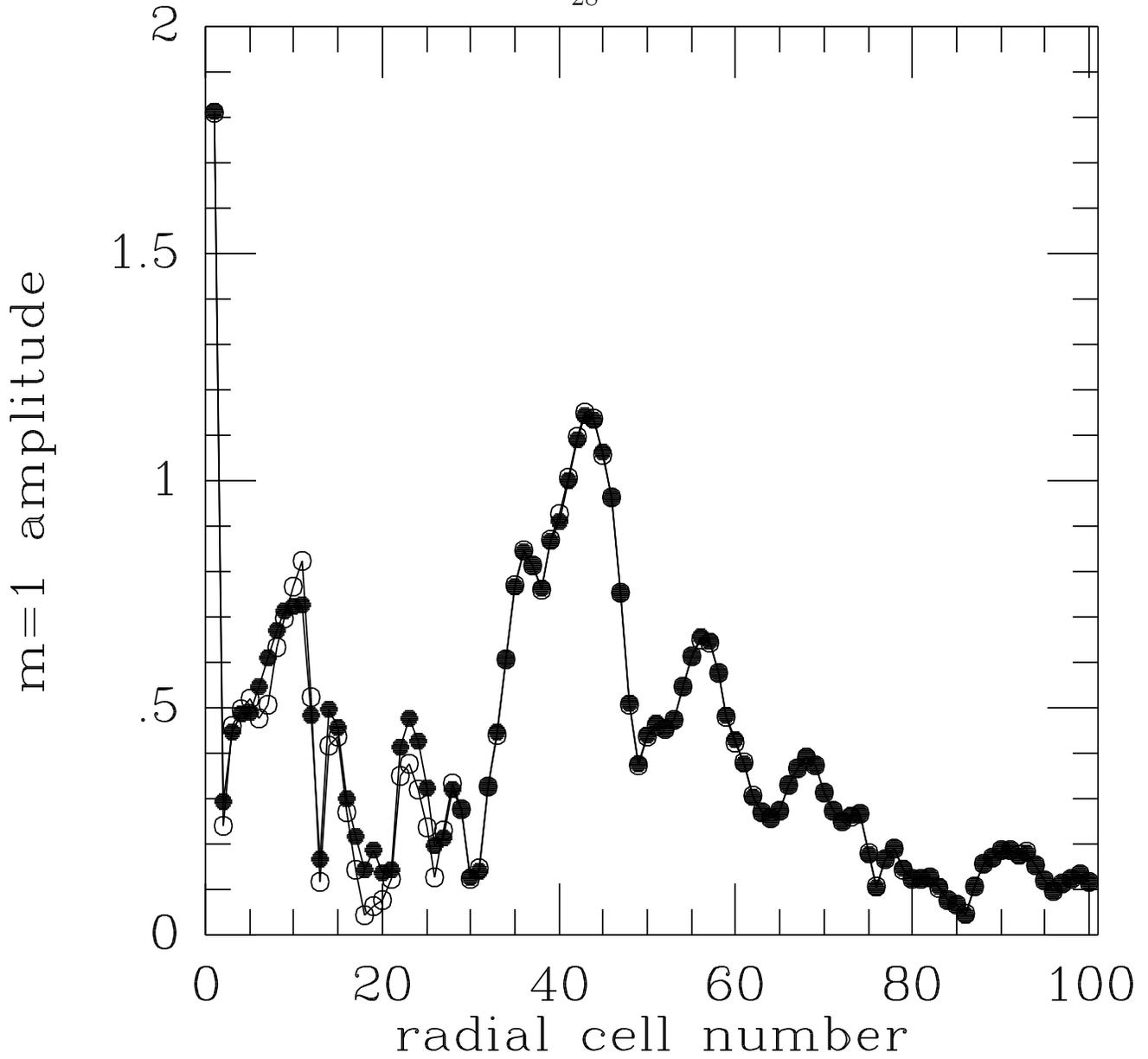}
\caption{Amplitudes of the $m = 1$ mode in a spherical harmonic expansion
of the midplane density distribution as a function of radial distance 
in the disk, with radial cell number 1 located at 4 AU and cell number
100 located at 20 AU. The amplitudes for model FL1 (filled symbols) 
and for model TE (open symbols) are shown at 328.2 and 328.3 yrs,
respectively.}
\end{figure}

\clearpage

\begin{figure}
\vspace{-2.0in}
\plotone{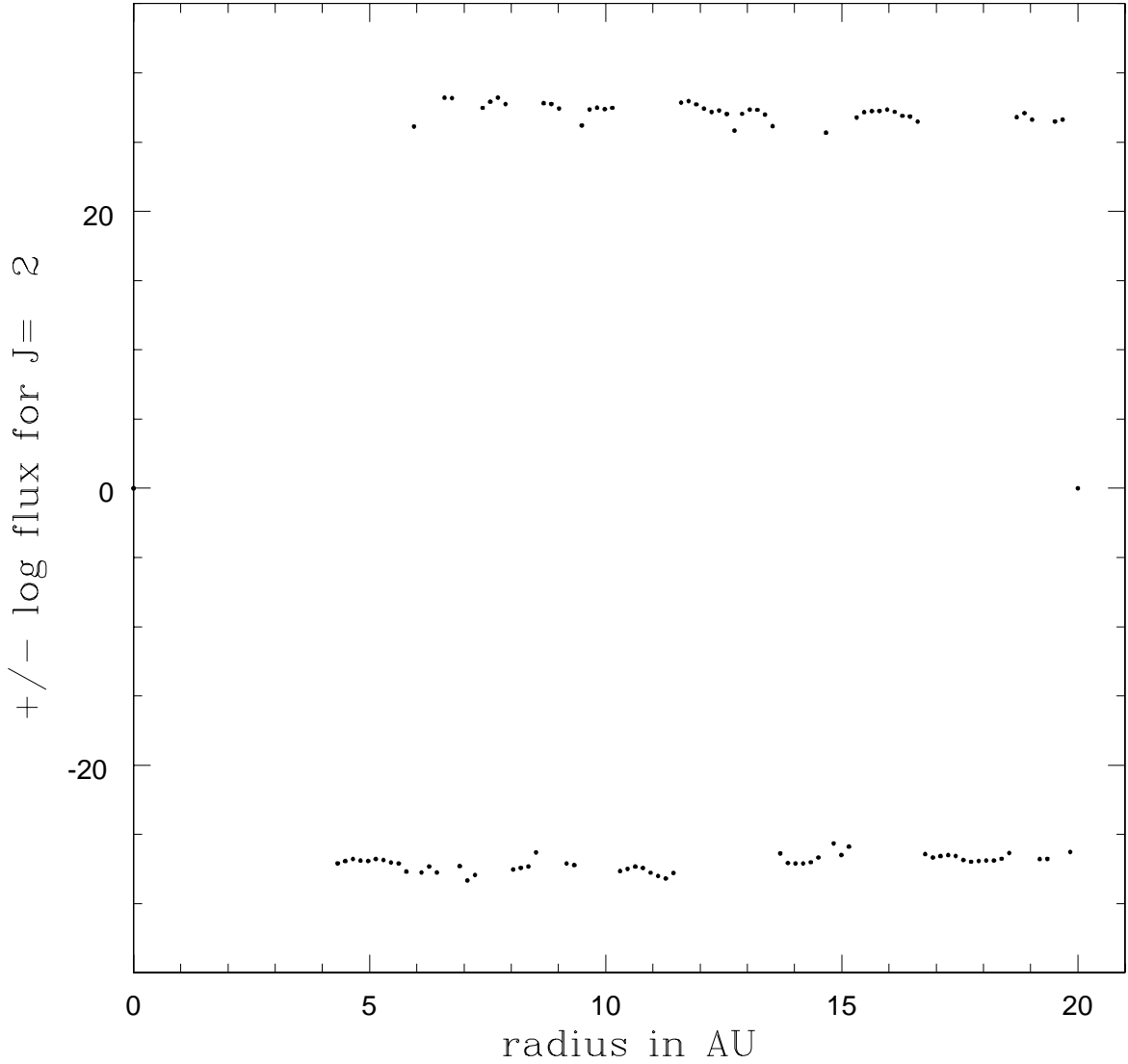}
\caption{Logarithm of the vertical convective flux (cgs units) 
as a function of radial distance for model FL1 at 329.6 yrs.
Values are plotted for a conical surface 0.3 degrees above
the midplane, where the fluxes must vanish. Positive
fluxes refer to upward transport, while negative fluxes
correspond to downward transport.}
\end{figure}

\clearpage

\begin{figure}
\vspace{-2.0in}
\plotone{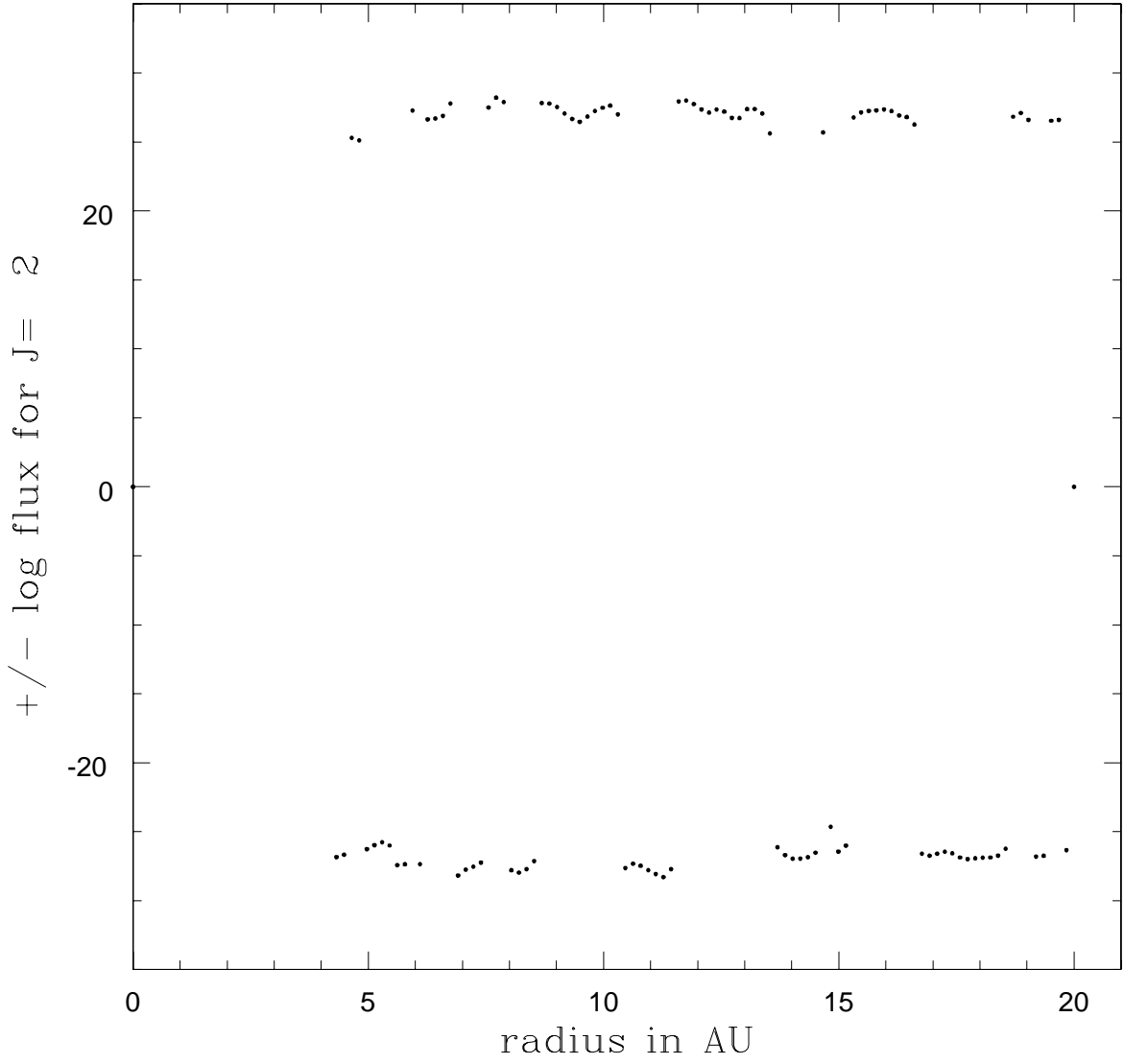}
\caption{Same as Figure 14, but for model TE at 330.3 yrs.}
\end{figure}

\end{document}